\documentclass{mhd}     

\usepackage{graphicx}
\usepackage{dcolumn}
\usepackage{bm}

\newcommand\Rey{\mbox{\textit{Re}}}  

\newcommand{\no}[1]{}

\mhdhead{0}{0}{1}   

\title{Energy dissipation rates in low-$\Rey_m$ MHD turbulence
with mean shear: Results for channel flow with spanwise  field }

\author{D. Krasnov\inst{1}, O. Zikanov\inst{2}, J. Schumacher\inst{1} and T. Boeck\inst{1}}
\institute{Fakult\"at Maschinenbau, Technische Universit\"at Ilmenau,\\
Postfach 100565, 98684 Ilmenau, Germany
\and
University of Michigan-Dearborn, 48128 MI, USA
}

\begin{document}

\maketitle
\begin{abstract}%
We examine the changes in kinetic energy dissipation of a 
turbulent channel flow caused by a spanwise magnetic field.
The numerical study is based on our  simulation data
from \cite{Krasnov:2008}
obtained  by  direct and large eddy simulations. 
We find that the Joule dissipation can exceed the viscous dissipation in the weakly
dissipative bulk region, but remains comparatively small in the
turbulence-generating near-wall region.
\end{abstract}

\section*{Introduction}
This paper continues our recent analysis \cite{Krasnov:2008} of 
turbulent channel flow in a  steady magnetic field imposed in the
spanwise (parallel to the walls but normal to the mean flow)
direction.  The magnetic Reynolds number is assumed small, and the
problem is solved in the framework of the quasi-static approximation
\cite{Roberts:1967}.  We are interested in this problem because  
the channel flow with spanwise magnetic field represents the simplest natural
flow configuration, in which we can study  turbulence under the
combination of two conditions common in industrial and laboratory
flows of liquid metals: imposed magnetic field and  mean shear.
At the same time, the transformation of  turbulent fluctuations
occurs without  direct interaction between the mean flow and the
magnetic field and without the dominating influence of  Hartmann
boundary layers.

A thorough study of the flow at high hydrodynamic Reynolds number
$\Rey\equiv UL/\nu=$ and moderate Hartmann number $Ha\equiv B L
(\sigma/\rho \nu)^{1/2}$ was conducted in \cite{Krasnov:2008}. Here,
$U$ and $L$ are the typical velocity and length scales, chosen in
our case as the centerline velocity of the laminar flow and half the
channel width, $B$ is the strength of the applied magnetic field,
and $\sigma$, $\rho$ and $\nu$ are the electric conductivity,
density, and kinematic viscosity of the fluid.  The results
confirmed and extended the conclusions of the earlier experimental
work \cite{Votsish:Kolesnikov:1977} and  numerical simulations
at low Reynolds number \cite{Lee:Choi:2001}. 
 We found that the main effect of the
magnetic field is the suppression of turbulent fluctuations,
particularly pronounced in the wall buffer regions, where the
fluctuations are generated by  strong mean shear.  The resulting
reduction of the turbulent momentum transport in the wall-normal
direction  decreases  the wall friction drag and 
transforms  the mean flow profile, which becomes steeper,
acquires higher centerline velocity, and loses the zone of the
logarithmic layer behavior.

Analyzing the anisotropy of turbulent fluctuations, we found
that the elongation of flow structures in the direction of the
magnetic field observed in  homogeneous zero-shear turbulence
(see, e.g. \cite{Zikanov:1998,Knaepen1:2004,Vorobev:2005}) appears
only in a weak form and only in the central area of the channel.  It
should be stressed that all the results of \cite{Krasnov:2008} were
obtained at moderate strength of the magnetic field, at $Ha\le 30$
for $\Rey=10^4$ and $Ha\le 40$ for $\Rey=2\times 10^4$.  Stronger
magnetic fields lead to an interesting intermittent flow regime, in which
turbulent states alternate with periods of nearly laminar behavior
\cite{Boeck:2008}.

The combination of the two factors, the mean shear and the imposed
magnetic field, also renders the channel flow an excellent
configuration for testing  computational models of turbulence,
in particular  LES models.  This was done in \cite{Krasnov:2008}
using the \emph{a-posteriori} comparison with the results of
high-resolution DNS.  In agreement with the earlier studies, such as
\cite{Knaepen1:2004,Vorobev:2008,Sarris:Kassinos:Carati:2007}, the
dynamic Smagorinsky model was shown to accurately reproduce the MHD
flow transformation.

In the present paper, we extend the investigation by considering the
viscous and magnetic (Joule) dissipations.  The databases generated
in the LES and high-resolution DNS computations \cite{Krasnov:2008}
are employed to analyze the effect of the magnetic field on the
dissipation rates in various zones of the flow.  An important goal
of the study is to interpret the observed flow transformation in
terms of the local energy balance.  We  provide further
verification of the dynamic Smagorinsky model by analyzing the
accuracy, with which the model reproduces the dissipation rates.
Such accuracy is generally considered one of the most important
characteristics of an LES model's performance.

\section*{Equations and numerical method}
We consider the flow of an incompressible, electrically conducting
fluid in a plane channel between two insulating walls located at $z=\pm L$. 
The flow is driven by a streamwise  pressure gradient 
 and submitted to a constant spanwise magnetic field
$\bm{B} = B \bm{e}_y$. The mean flow velocity $U_q$ is kept constant in the
simulations. The dimensional governing
equations and boundary conditions for the velocity field $u_i$, 
pressure $p$, electric current density $J_i$ and electric 
potential $\phi$ are
\begin{eqnarray}
\label{MHDeqn} \hskip-15mm& & \frac{\partial u_i}{\partial t} +
u_j\frac{\partial u_i}{\partial x_j} = -\frac{1}{\rho}\frac{\partial
p}{\partial x_i}  +  \frac{1}{\rho}\frac{\partial}{\partial
x_j}\left(\tau_{ij}+\tau^a_{ij}\right) + \frac{1}{\rho}\epsilon_{ijk}J_j B_k,\no{MHDeqn}\\
 \hskip-15mm& &  J_i =  \sigma\left(-\frac{\partial \phi}{\partial x_i} +  \epsilon_{ijk}u_j B_k\right),\\
\hskip-15mm& & \frac{\partial u_i}{\partial x_i} =  
\frac{\partial J_i}{\partial x_i} =  0,\\
\label{MHDeqn_bc} \hskip-15mm& & u = v = w = \frac{\partial
\phi}{\partial z}=0 \hskip3mm\mbox{at } z=\pm L.\no{MHDeqn_bc}
\end{eqnarray}
In these equations,
$\tau_{ij}=2\rho\nu S_{ij}$ is the resolved viscous stress tensor, where $2S_{ij}=
\partial u_i/\partial x_j+\partial u_j/\partial x_i$ is the rate-of-strain
tensor. The additional subgrid stress tensor $\tau^a_{ij}$, which appears in
 LES  computations, is given by $\tau^a_{ij}=2 \rho \nu_T
\bar{S}_{ij}$. It is computed using the local filtered rate of strain
tensor $\bar{S}_{ij}$, with the eddy viscosity $\nu_T$ determined by
the dynamic Smagorinsky model 
with averaging in  horizontal planes.

Variables are made non-dimensional by  the laminar
centerline velocity $U$, half-channel width $L$ and $ULB$ (for $\phi$).
The non-dimensional parameters are  $\Rey$ and  $Ha$. 
The problem is solved numerically in a domain with the periodicity lengths 
$L_x/L=2\pi$ and $L_y/L=\pi$. 
The pseudo-spectral  numerical method is described in
\cite{Krasnov:2008,Krasnov:2004}.  
The numerical resolution is  $N_x\times N_y\times
N_z= 256^3$ collocation points in the simulations with
$\Rey=10^4$ and $512^2\times 256$ collocation points at $\Rey=2\times
10^4$.
The spatially filtered equations for LES are solved using $64^3$
and $128^3$ collocation points at $\Rey=10^4$ and 
$\Rey=2\times 10^4$, respectively.  

The equation of the local kinetic energy balance is formed from the
momentum equation (\ref{MHDeqn}) by multiplying it by the velocity
vector $\bm{u}$, making  use of Ohm's law, and rearranging such
that the energy transport terms are entirely separated from the
local dissipation terms.  The result is the equation 
\begin{equation}\label{energy}
\frac{\partial e}{\partial t}=-\nabla \cdot
\bm{N}-\varepsilon_{\textrm{visc}}-\varepsilon_{\mu}-\varepsilon_{\textrm{sgs}},
\end{equation}
where $e=u^2/2$ is the specific kinetic energy,
\begin{equation}\label{visc}
\varepsilon_{\textrm{visc}}\equiv \tau_{ij}S_{ij}/\rho
\end{equation}
is the viscous dissipation rate, and
\begin{equation}\label{joule}
\varepsilon_{\mu}\equiv J^2/\sigma\rho
\end{equation}
is the Joule dissipation rate. The additional 
dissipation rate due to the subgrid
stresses is
\begin{equation}\label{sgs}
\varepsilon_{\textrm{sgs}} \equiv \tau^a_{ij} \bar{S}_{ij}/\rho.
\end{equation}
$\bm{N}$ is the energy flux vector
due to convective transport, pressure force, friction force, subgrid stresses, and
Lorentz force. It is given by 
\begin{equation}\label{flux}
\rho N_i=\rho u_i \left(\frac{u^2}{2}+\frac{p}{\rho}
\right)-u_i\tau_{ij}-u_i\tau^a_{ij} -\phi J_i.
\end{equation}
In the case of DNS, the subgrid stress contribution is absent from all
of the above equations.

\section*{Results} 

\begin{figure}[htb]
\scriptsize{
\centerline{
\includegraphics[width=0.48\textwidth]{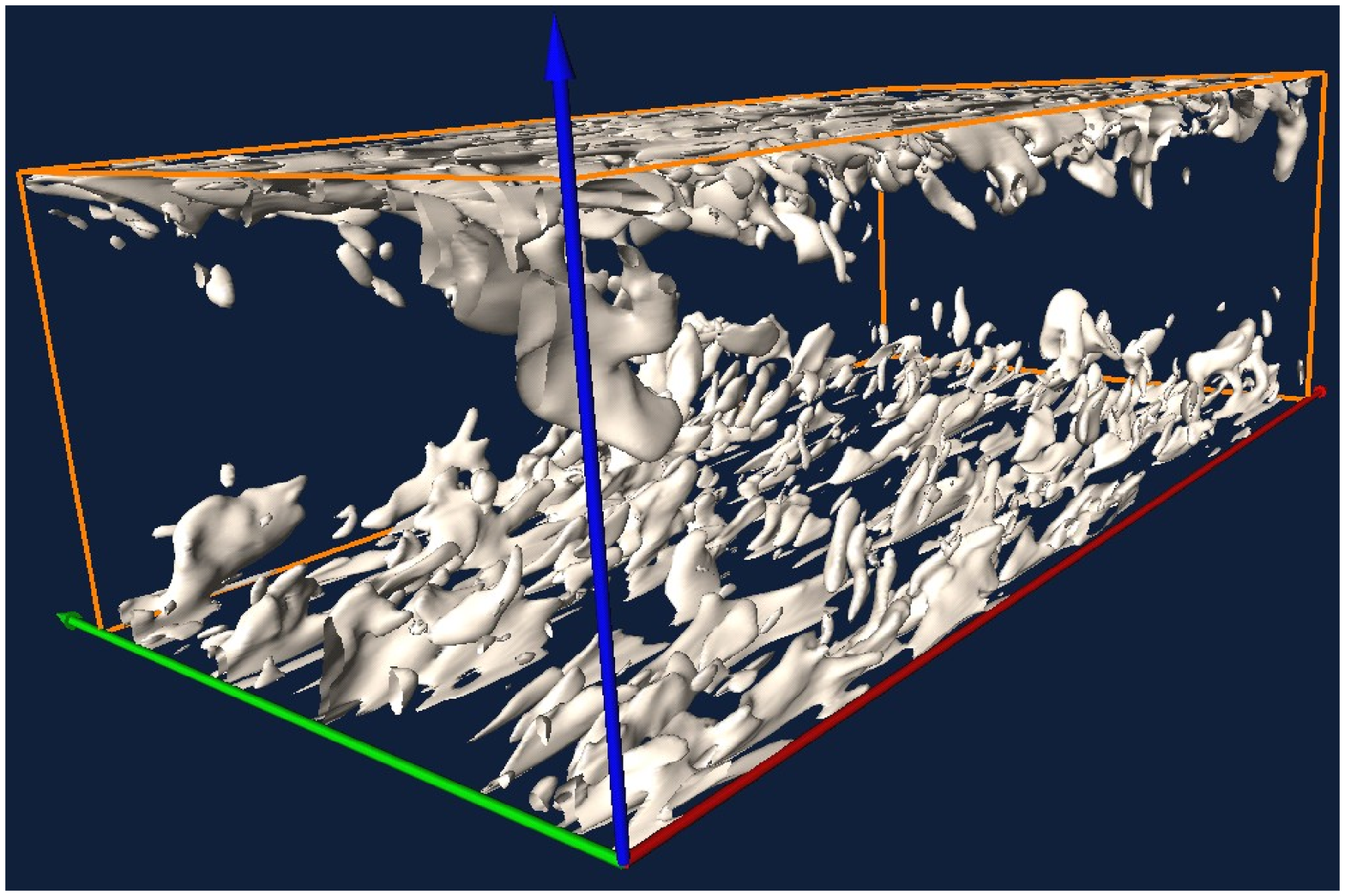}
\includegraphics[width=0.48\textwidth]{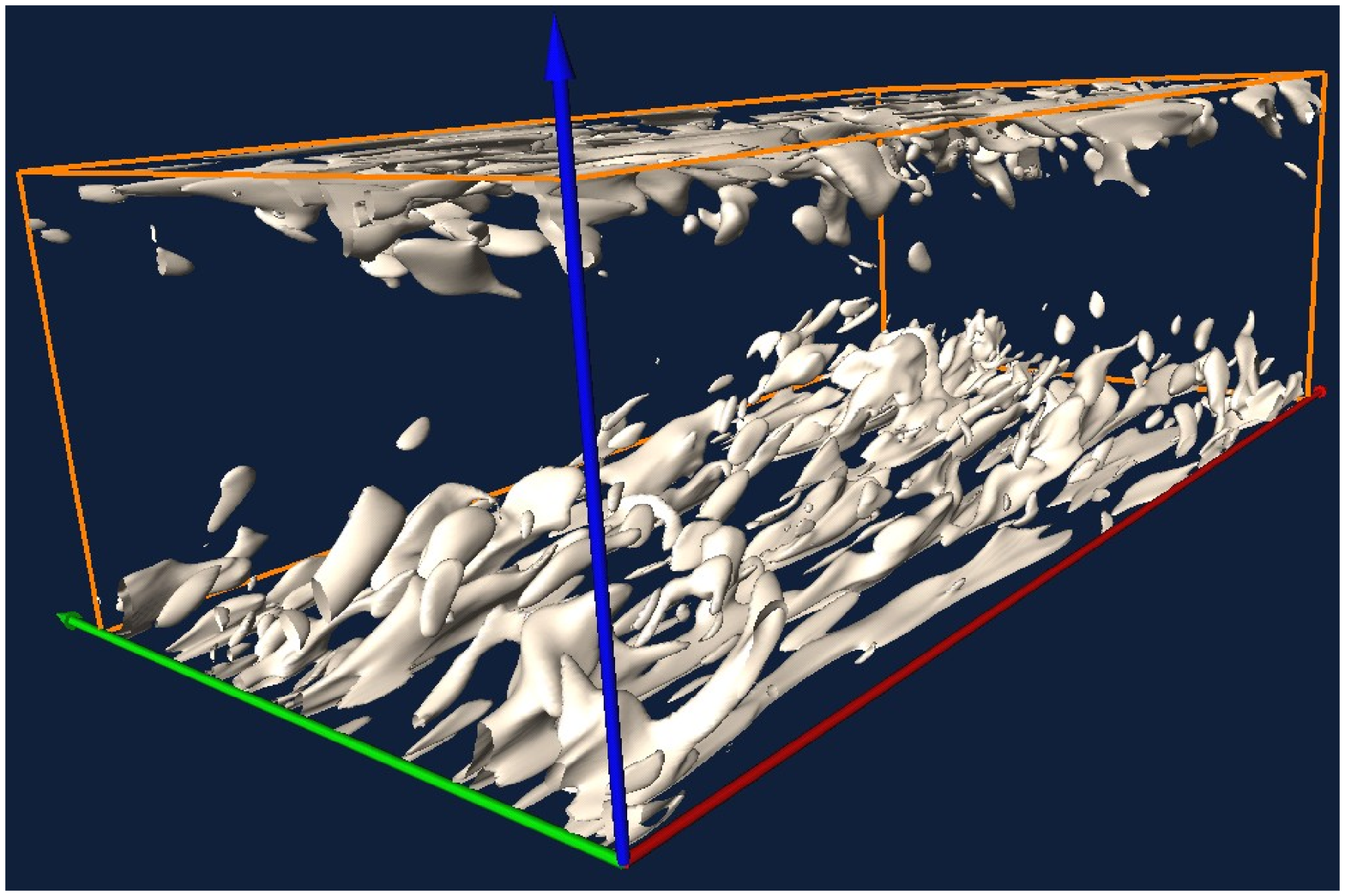}
}
}
\caption{Iso-surfaces $\varepsilon_{\mu}/\varepsilon_{\mu}^{rms} = 2$
of the Joule dissipation rate 
for $Ha=10$ (left) and $Ha=30$ (right) from DNS with $\Rey=10000$.
Snapshots of the entire computational domain are shown.}
\label{fig3:re10_joule}
\end{figure}

We first consider the spatial structure of the Joule dissipation field.
Fig. \ref{fig3:re10_joule} 
illustrates the trend observed in our earlier paper \cite{Krasnov:2008} 
for the velocity field:
a stronger magnetic field leads to turbulence suppression and smoother, larger
structures.

\begin{figure}[h]
\scriptsize{
\parbox{0.55\linewidth}{(a)}\parbox{0.40\linewidth}{(b)}
\centerline{
\includegraphics[width=0.49\textwidth]{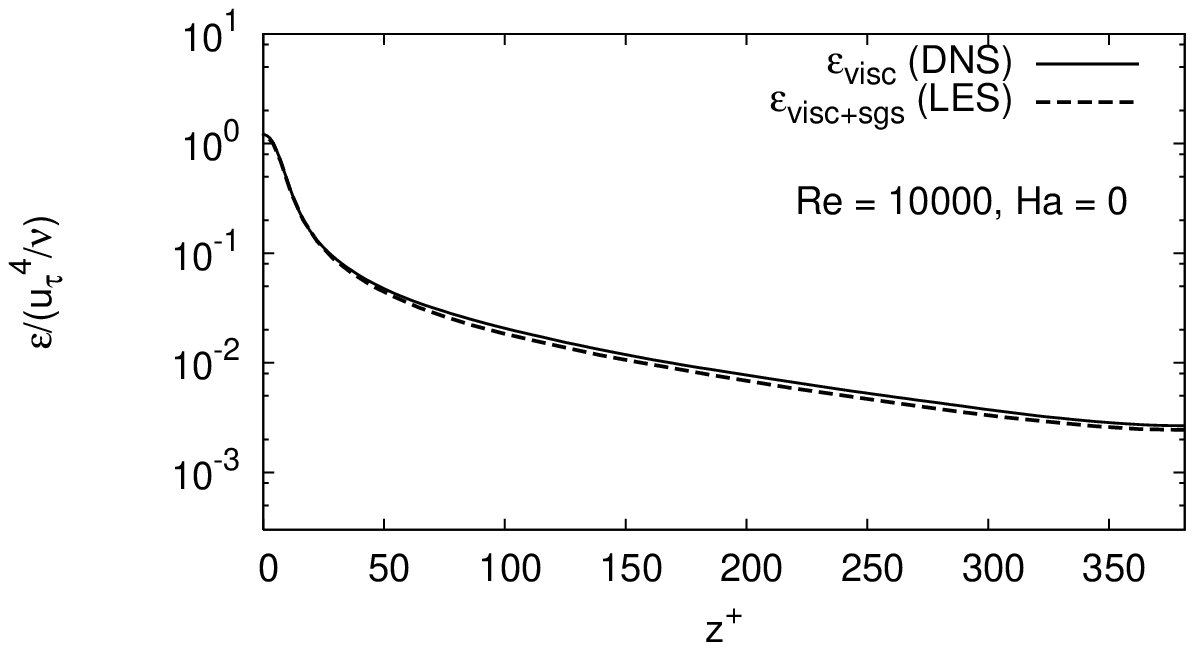}
\includegraphics[width=0.49\textwidth]{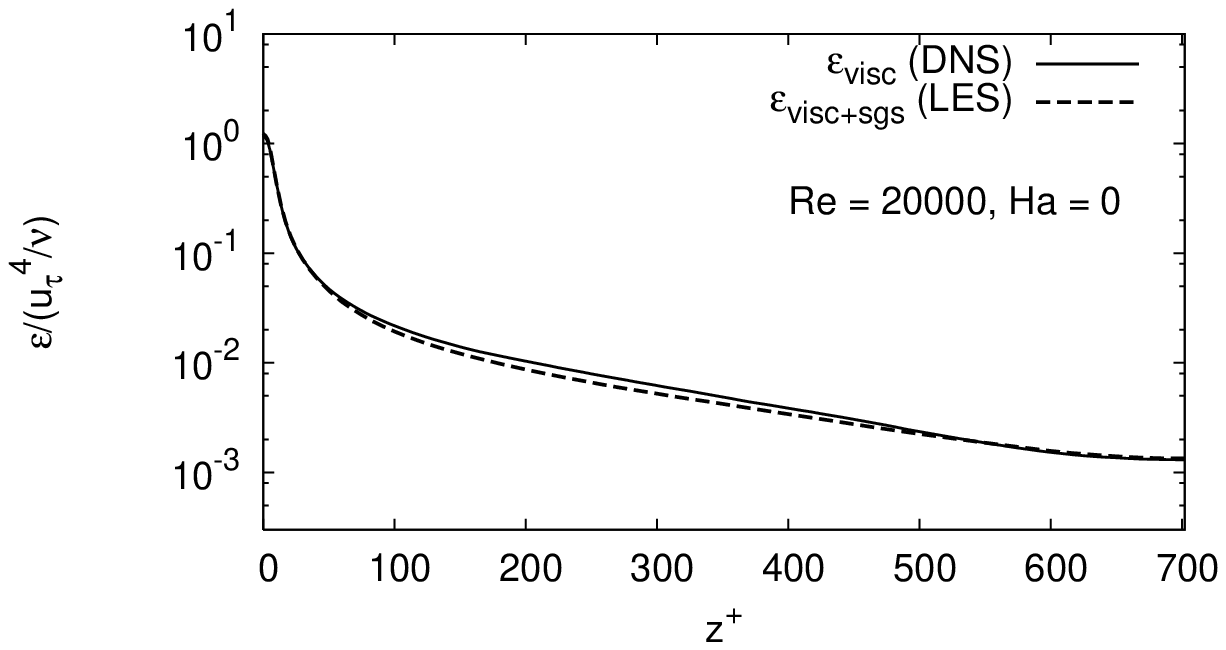}
}
} 
\caption{Time-averaged profiles of $\varepsilon_{\textrm{visc}}$ 
for \emph{(a)} $Re=10000$ at $Ha = 0$ and \emph{(b)}
$Re=20000$ at $Ha = 0$. In case of LES the total viscous dissipation
$\varepsilon_{\textrm{visc+sgs}}$ is plotted. 
Normalization is by $u^4_{\tau}/\nu$
obtained for each case from DNS. The distance $z^+$ from the wall is
measured in units of the friction length $\nu/u_{\tau}$.}
\label{fig1a:dns_les}
\end{figure}

\begin{figure}[htb]
\scriptsize{
\parbox{0.55\linewidth}{(a)}\parbox{0.40\linewidth}{(b)}
\centerline{
\includegraphics[width=0.49\textwidth]{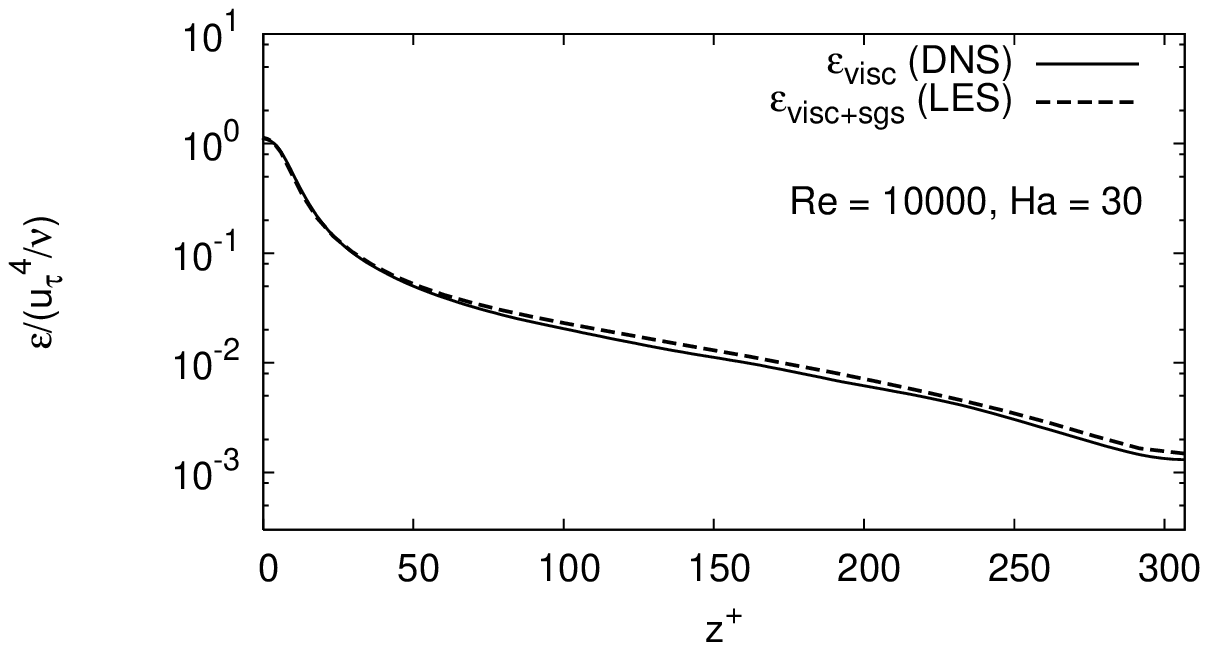}
\includegraphics[width=0.49\textwidth]{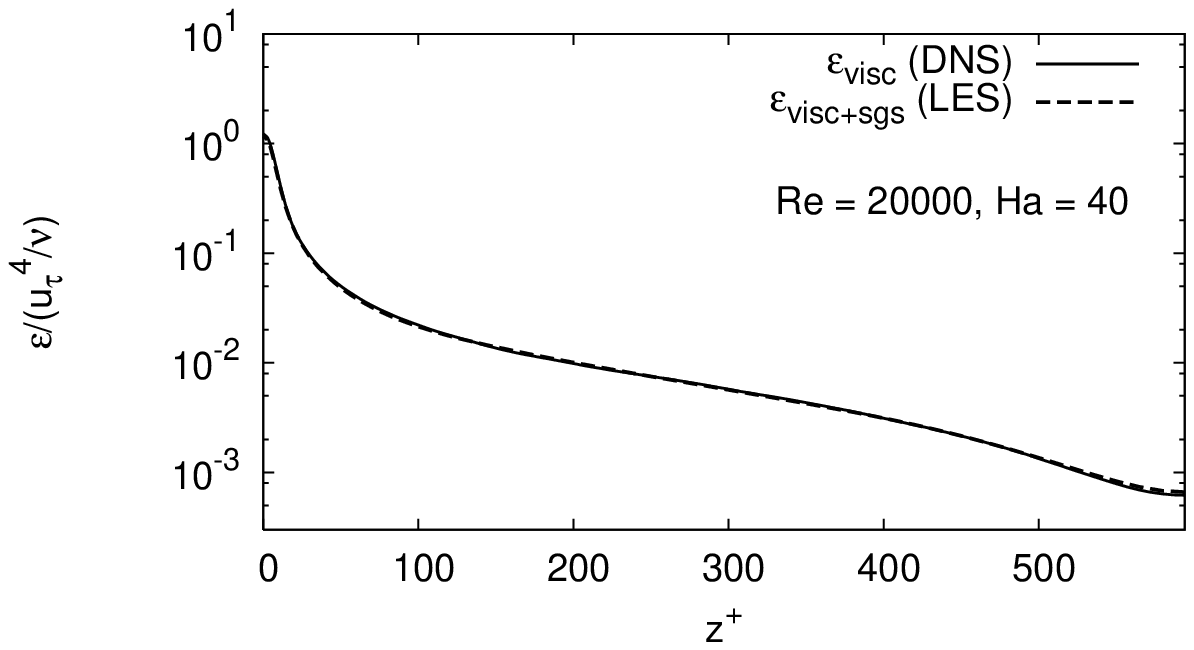}
}
\parbox{0.55\linewidth}{(c)}\parbox{0.40\linewidth}{(d)}
\centerline{
\includegraphics[width=0.49\textwidth]{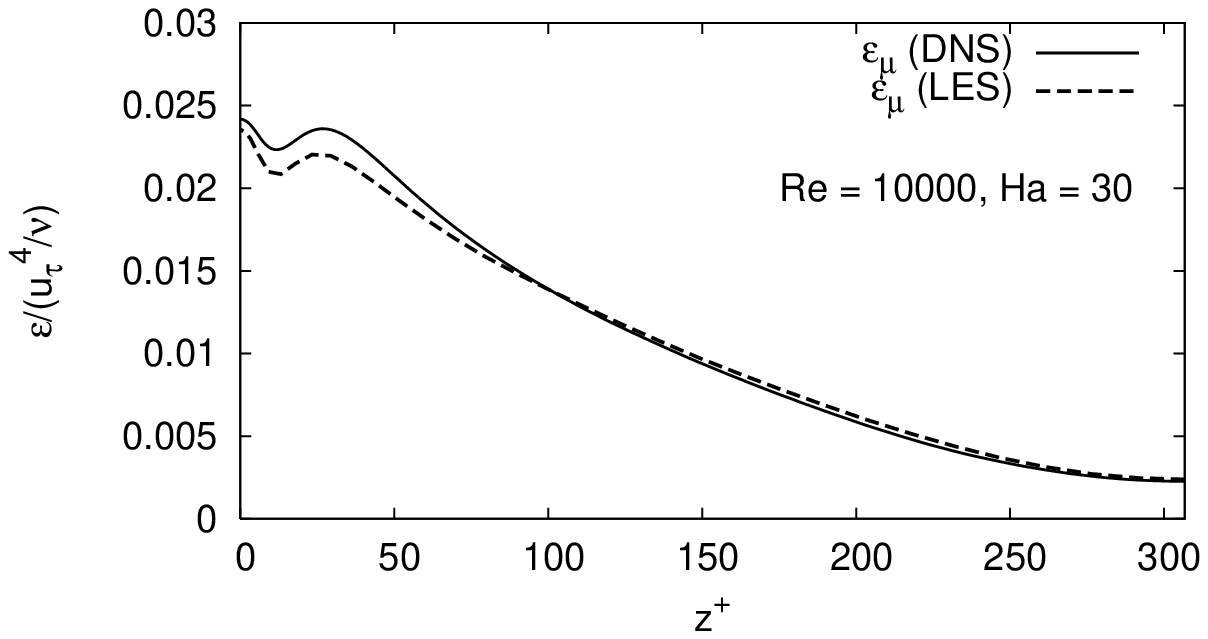}
\includegraphics[width=0.49\textwidth]{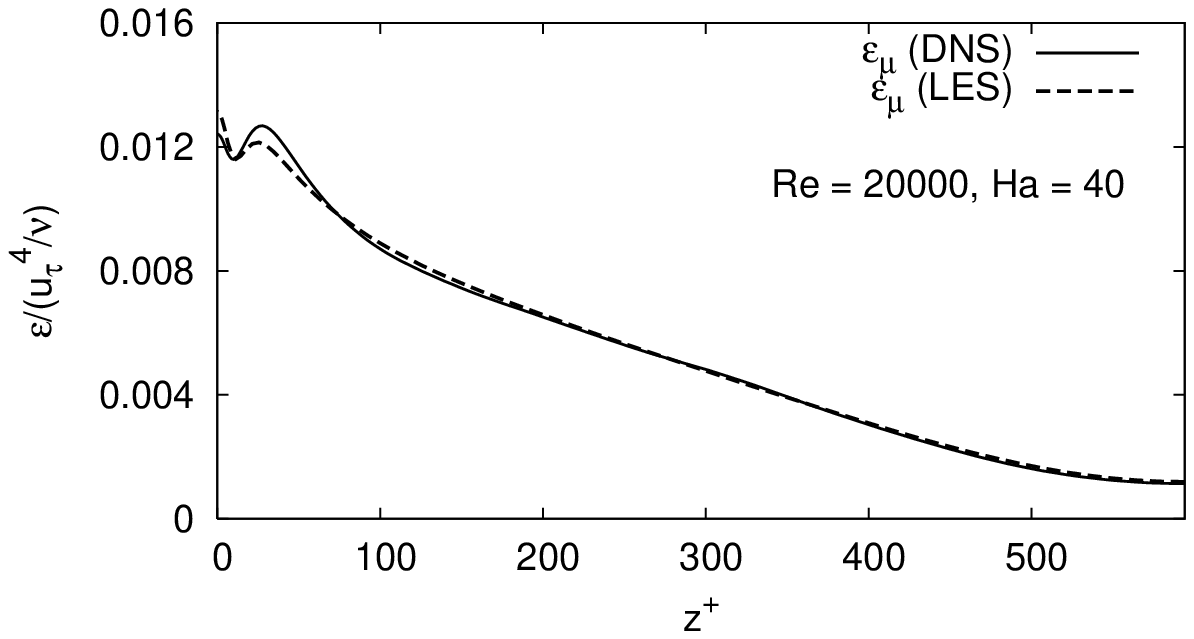}
}
} 
\caption{Time-averaged profiles of  $\varepsilon_{\textrm{visc}}$ and 
$\varepsilon_{\mu}$   for \emph{(a,c)}
$Re=10000$ at $Ha = 30$ and \emph{(b,d)} $Re=20000$ at $Ha = 40$.
In case of LES the total viscous dissipation 
$\varepsilon_{\textrm{visc+sgs}}$~
and the resolved part of the Joule dissipation
$\varepsilon_{\mu}^{\textrm{resolved}}$
are plotted. Normalization is by $u^4_{\tau}/\nu$ obtained for
each case from DNS.}
\label{fig1b:dns_les}
\end{figure}

For  further analysis,  we focus on 
the average distributions of dissipation rates and their
transformation by the magnetic field.
The averaging is performed in  $x$-$y$-planes and in time over several
convective times $L/U$.
We begin with the demonstration that LES reproduces the dissipation rates
with satisfactory accuracy.
Fig. \ref{fig1a:dns_les} shows close agreement between the 
mean profiles of the viscous dissipation rate  from DNS and the sum
of viscous and sub-grid scale dissipation rates from LES in the 
 non-magnetic cases. Figs. \ref{fig1b:dns_les}(a,b) show similar
 agreement for the cases of the
 strongest magnetic fields realized in the simulations.
\begin{table}[t]
\begin{center}
\begin{tabular}{@{}cccc|cccc@{}}
\hline
{Run} & {$Ha$} & {$\varepsilon_{\mu}/\varepsilon_{\textrm{tot}}$} & {$\varepsilon_{\textrm{sgs}}/\varepsilon_{\textrm{tot}}$} &
{Run} & {$Ha$} & {$\varepsilon_{\mu}/\varepsilon_{\textrm{tot}}$} & {$\varepsilon_{\textrm{sgs}}/\varepsilon_{\textrm{tot}}$}\\[6pt]

\hline \multicolumn{8}{c}{$Re=10000$ \hskip30mm $Re=20000$} \\ \hline

 LES & $10$ & $0.020$ & $0.116$ & LES & $20$ & $0.042$ & $0.113$ \\
\hline
 LES & $20$ & $0.077$ & $0.095$ & LES & $30$ & $0.088$ & $0.099$ \\
\hline
 DNS & $30$ & $0.158$ & $0.0$ & DNS & $40$ & $0.142$ & $0.0$ \\
 LES & $30$ & $0.154$ & $0.067$ & LES & $40$ & $0.147$ & $0.080$ \\
\hline
\end{tabular}
\end{center}
\caption{Time- and volume-averaged subgrid-scale and Joule dissipation 
rates relative to the total energy dissipation rate
 $\varepsilon_{\textrm{tot}}=\varepsilon_{\textrm{visc}}+\varepsilon_{\mu}+
\varepsilon_{\textrm{sgs}}$.}
\label{table:eps_frac}
\end{table}
For the latter two cases, the Joule dissipation rates 
 shown in Figs. \ref{fig1b:dns_les}(c,d) also show good agreement. It should be
 mentioned that the comparison should, in principle, be done between the LES and
 filtered DNS results. Filtering was not performed in our analysis. Unlike
the viscous dissipation, the Joule dissipation rate varies with the length scale 
in about the same way as the kinetic energy, and   the Joule dissipation 
in the subgrid scales should therefore be quite small. The slightly larger
discrepancies in $\varepsilon_{\mu}$  near the wall at $\Rey=10^4$ 
in Fig. \ref{fig1b:dns_les}(c) are most likely caused by a comparatively coarser 
LES resolution than for $\Rey=2\times 10^4$ (Fig. \ref{fig1b:dns_les}(d)).
In any case, our LES results are quite satisfactory for the mean profiles
of the dissipation rates. This observation is further supported by 
Table \ref{table:eps_frac}, 
which shows that the fraction of the Joule dissipation relative to the 
total dissipation is well represented in the   LES model.

\begin{figure}[htb]
\scriptsize{
\centerline{
\includegraphics[width=0.75\textwidth]{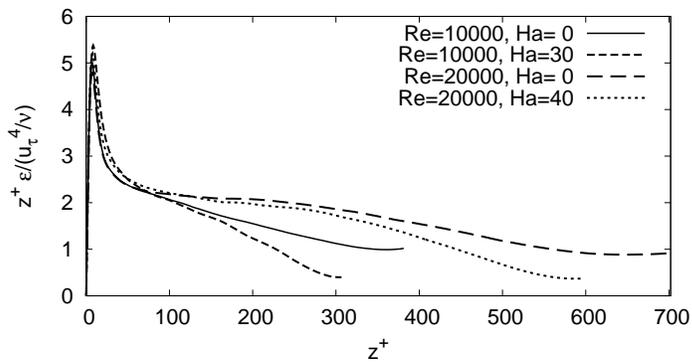}
}
} 
\caption{Compensated viscous dissipation 
$z^+ \varepsilon_{\textrm{visc}}$ from DNS 
for $\Rey=10000$ at $Ha = 0,30$ and $\Rey=20000$ at $Ha = 0,40$.
The plots are continued up to the middle of the channel, which
varies because of different friction lengths $\nu/u_{\tau}$.}
\label{fig1c:eps_visc}
\end{figure}

Non-magnetic turbulent channel flows are characterized by  local 
equilibrium between kinetic energy production and dissipation
and by the approximate relation  
 $\varepsilon_{\textrm{visc}}\sim 1/z^+$ 
in the logarithmic layer \cite{Pope:2000}.  The product
$z^+\varepsilon_{\textrm{visc}}$ should therefore be approximately constant.
Fig. \ref{fig1c:eps_visc} shows this quantity as found in our DNS computations.
We can identify
a  plateau for $\Rey=2\times 10^4$ in both magnetic and non-magnetic cases,
although it is shown in \cite{Krasnov:2008} that 
the mean velocity profile for $Ha=40$ deviates significantly from the
logarithmic law. For $\Rey=10^4$  the quantity 
$\varepsilon_{\textrm{visc}}$ decays somewhat faster than
$1/z^+$ in the non-magnetic as well as the magnetic case, and the  plateau is
therefore not observed. 

\begin{figure}[htb]
\scriptsize{
\parbox{0.55\linewidth}{(a)}\parbox{0.40\linewidth}{(b)}
\centerline{
\includegraphics[width=0.49\textwidth]{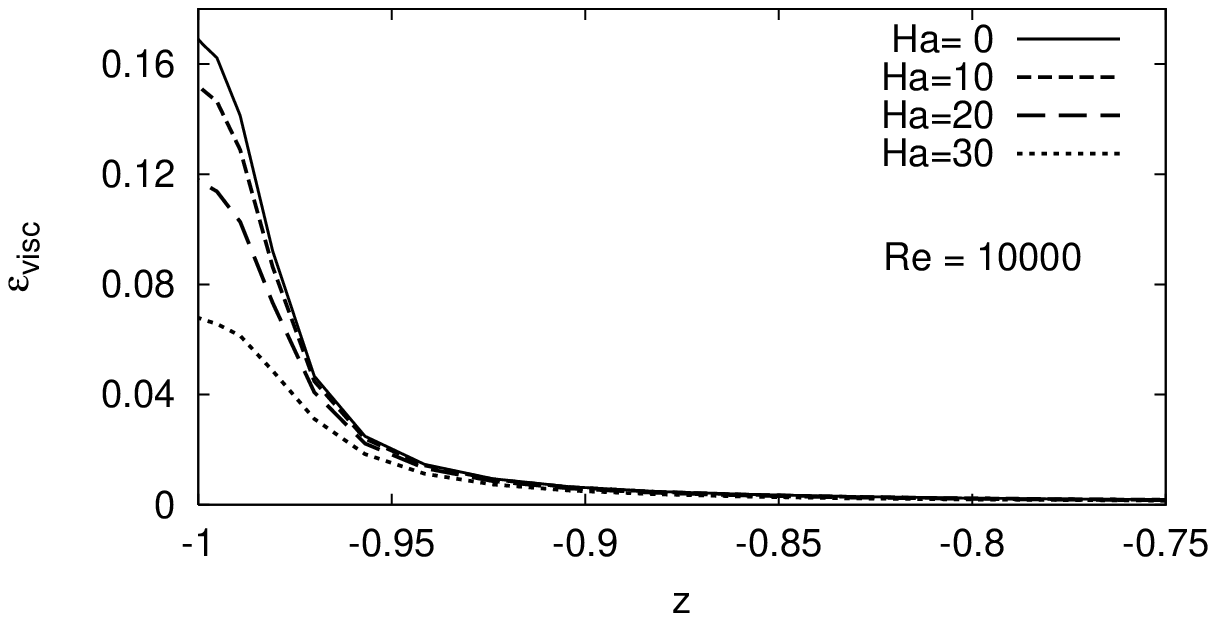}
\includegraphics[width=0.49\textwidth]{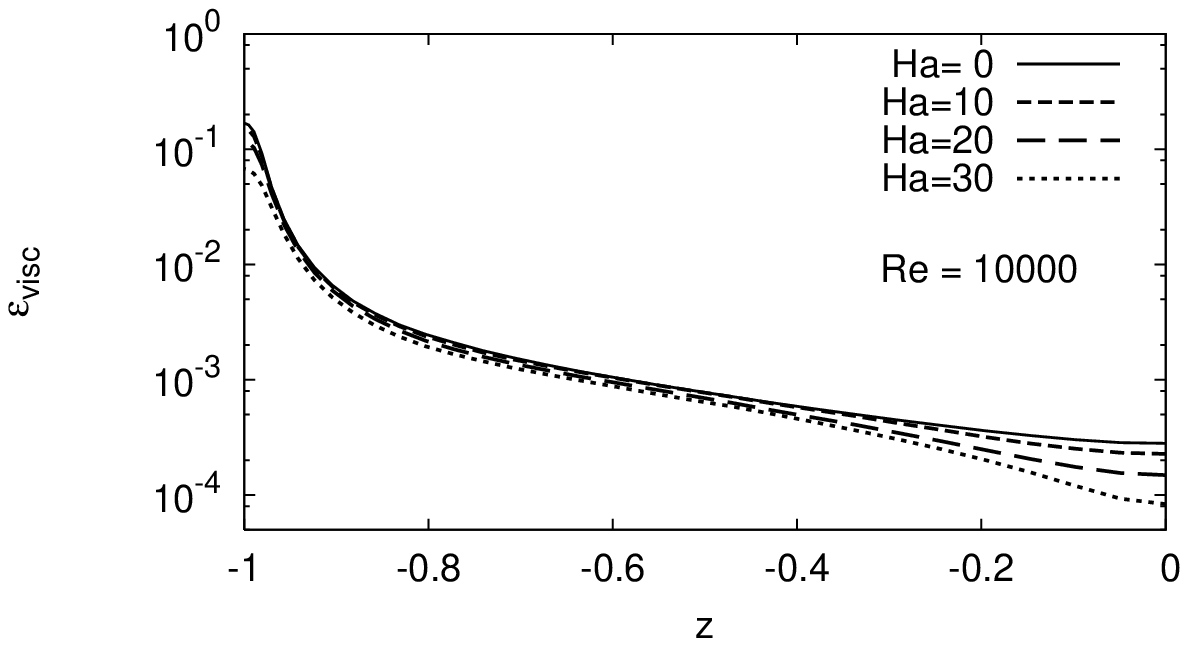}
}
\parbox{0.55\linewidth}{(c)}\parbox{0.40\linewidth}{(d)}
\centerline{
\includegraphics[width=0.49\textwidth]{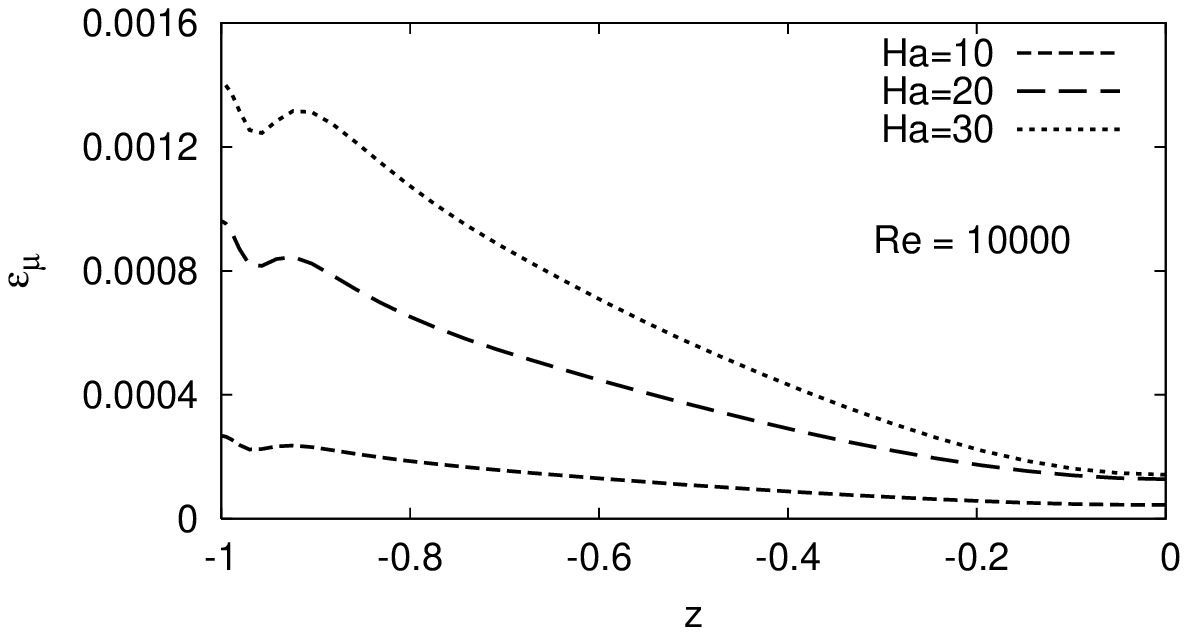}
\includegraphics[width=0.49\textwidth]{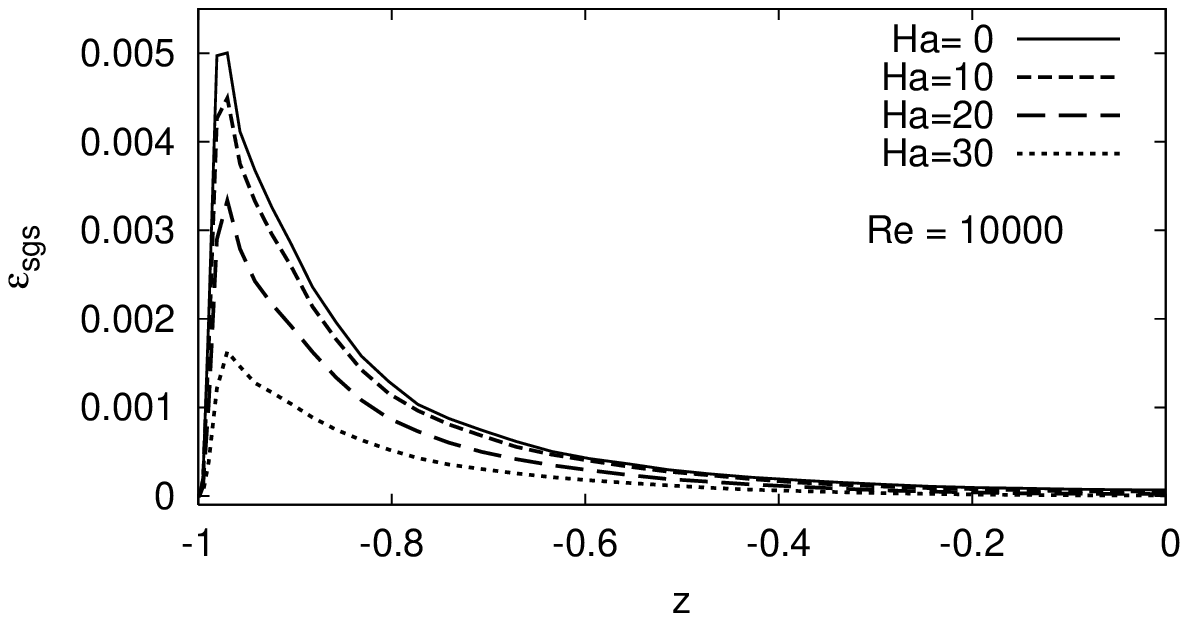}
}
} 
\caption{Horizontally and time-averaged
rates of \emph{(a,b)} viscous, \emph{(c)} Joule  and 
\emph{(d)} subgrid-scale 
dissipation rates from LES for different $Ha$ at $\Rey=10000$. 
The viscous term is shown for the near-wall region
\emph{(a)} and, to facilitate the comparison in the bulk region,
in log-scale  \emph{(b)}. 
Quantities are normalized
by $U^3_{q}/L$, which is identical for the different $Ha$.}
\label{fig2:les_vs_ha}
\end{figure}

The systematic trends in the behavior of the dissipation rates with increasing
magnetic field  are illustrated in Fig. \ref{fig2:les_vs_ha} by the LES data
obtained at $\Rey=10^4$. 
The results for $\Rey=2\times 10^4$ are qualitatively identical and,
therefore, omitted. Figs. \ref{fig2:les_vs_ha}(a,b) show that 
the viscous dissipation rate is significantly reduced by the magnetic field near
the wall and in the middle of the channel. The reduction at the walls
is expected because of the suppression of the intensity of the turbulent
fluctuations observed in  \cite{Krasnov:2008}.
The reduction in the middle is less significant for the overall level of
dissipation since this region contributes very little, but the relative
change in $\varepsilon_{\textrm{visc}}$ 
by a factor of about $3.3$   from $Ha=0$ to $Ha=30$  is larger than at
the wall. The Joule dissipation rate shown in Fig. \ref{fig2:les_vs_ha}(c) 
increases significantly 
with $Ha$ but without much change in the profile shape.
The subgrid-scale dissipation rate in Fig. \ref{fig2:les_vs_ha}(d) decreases 
significantly from $Ha=0$ to
$Ha=30$, as can be expected from the magnetic damping of the turbulence
and the corresponding depletion of the subgrid scales.

\begin{figure}[htb]
\scriptsize{
\parbox{0.55\linewidth}{(a)}\parbox{0.40\linewidth}{(b)}
\centerline{
\includegraphics[width=0.49\textwidth]{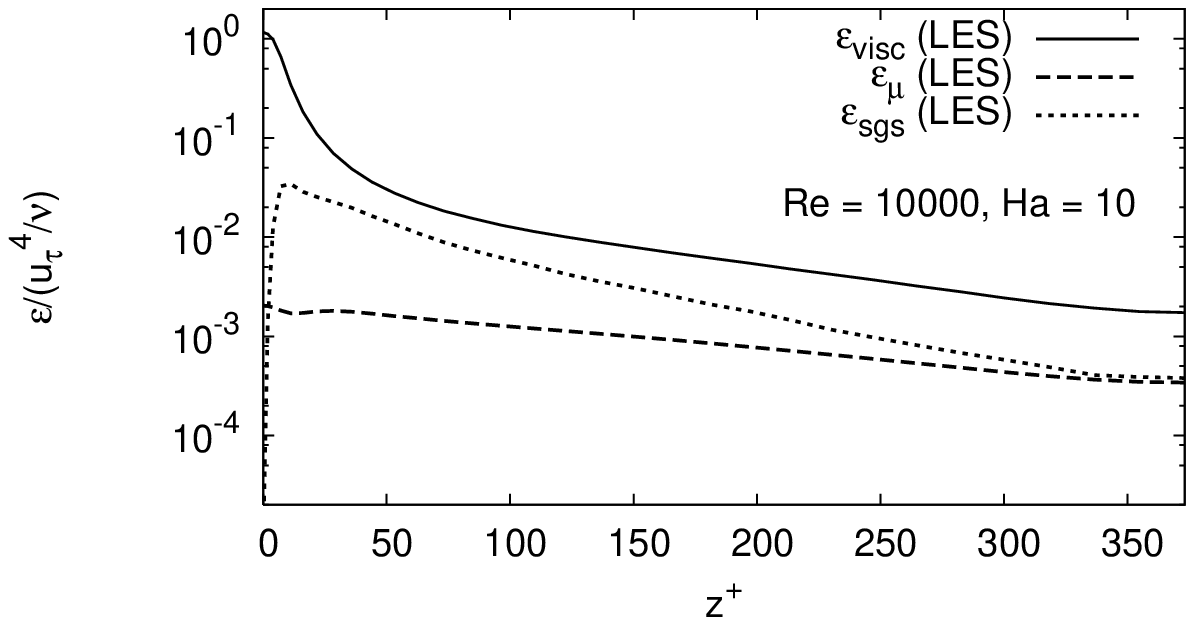}
\includegraphics[width=0.49\textwidth]{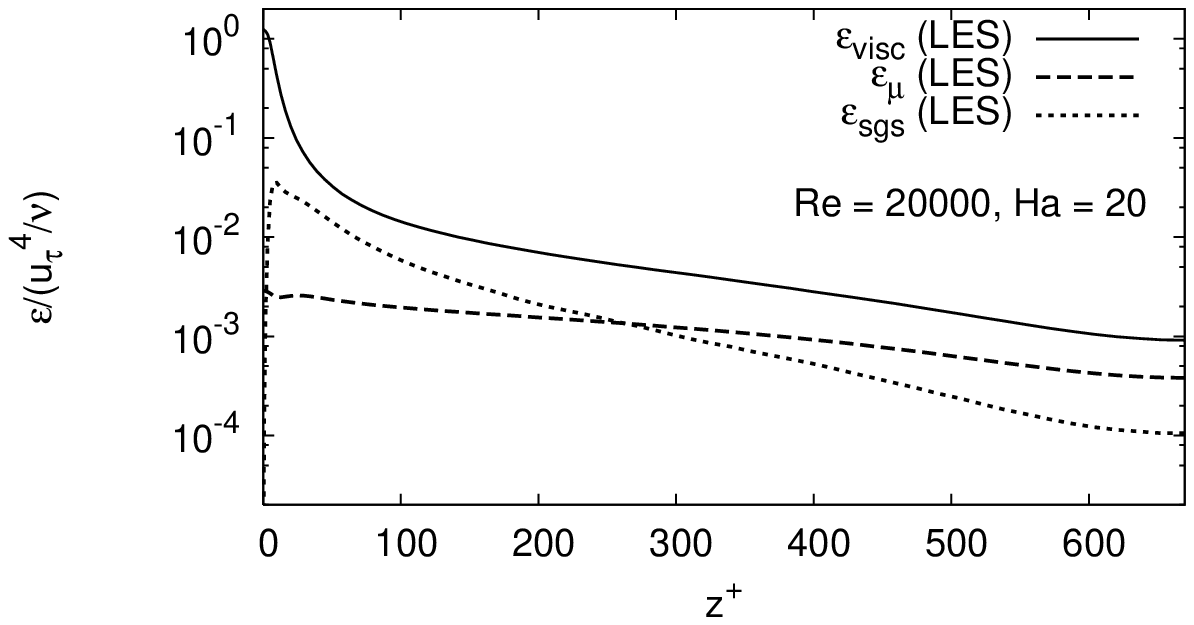}
}
\parbox{0.55\linewidth}{(c)}\parbox{0.40\linewidth}{(d)}
\centerline{
\includegraphics[width=0.49\textwidth]{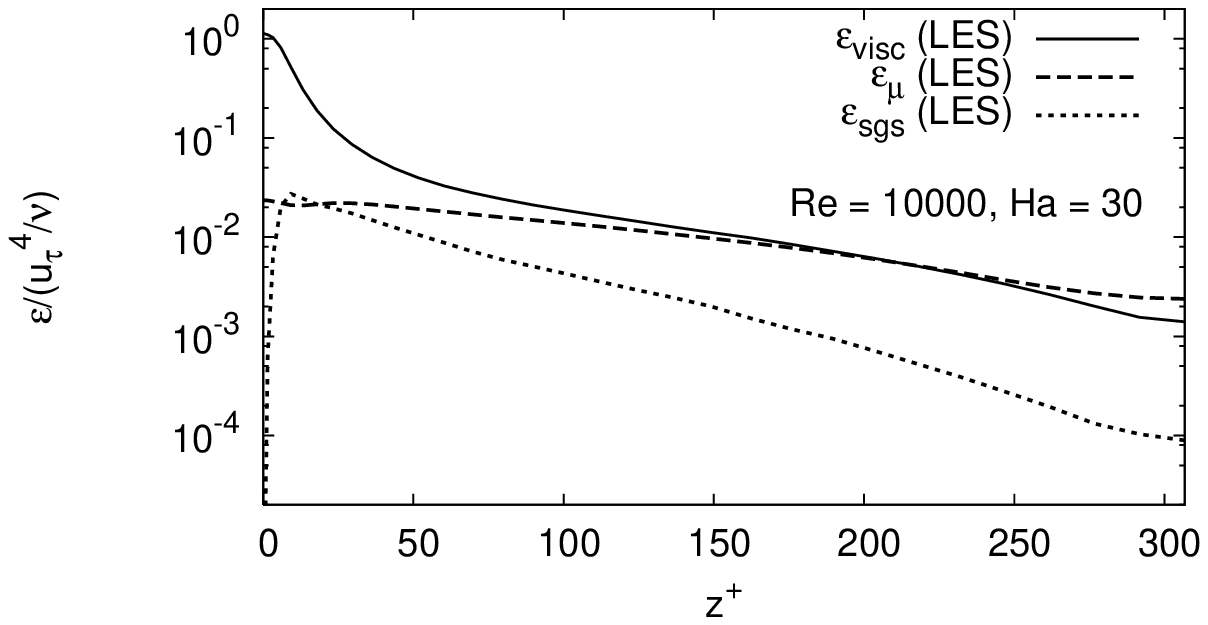}
\includegraphics[width=0.49\textwidth]{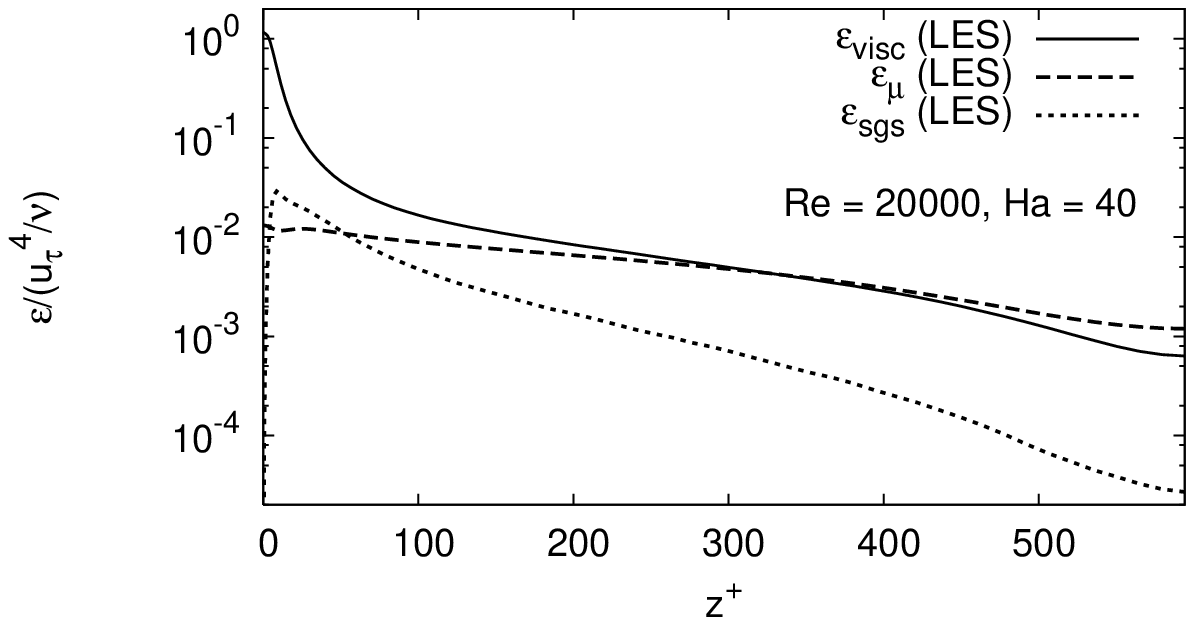}
}
} 
\caption{Mean profiles of the  dissipation rates obtained from 
LES  for $\Rey=10000$ (left) and $\Rey=20000$
(right) at different $Ha$.
Normalization
is by $u^4_{\tau}/\nu$ obtained for each case from DNS.
}
\label{fig2:mhd_vs_sgs}
\end{figure}

The relative contributions of the dissipation rates are
compared in Fig. \ref{fig2:mhd_vs_sgs}, 
which shows mean profiles calculated in LES for $\Rey=10^4$
and $\Rey=2\times 10^4$. At low $Ha$, the contribution from the 
Joule dissipation rate
is everywhere smaller than that of the viscous dissipation, 
but for the highest $Ha$, the Joule
dissipation becomes comparable or even larger than 
$\varepsilon_{\textrm{visc}}$  in the bulk region. 
The overall contributions are compared in Table \ref{table:eps_frac}.
The contribution of the subgrid-scale dissipation 
diminishes considerably in the bulk with increasing
$Ha$, especially for $\Rey=2\times 10^4$ with its larger number of grid points.
The overall contribution of the subgrid-scale dissipation remains nonetheless
of the order of $10\%$.


In conclusion we remark that both the Joule and viscous
dissipation rates peak at the wall. The Joule dissipation rate 
attains a local maximum in the buffer region, where the turbulence is primarily
generated.
The shape of the Joule dissipation profile is not much affected by the strength of the
magnetic field. Its contribution to the total dissipation remains fairly modest,
but it  can exceed the viscous dissipation locally (in the bulk) before the
intermittent dynamics described in \cite{Boeck:2008} appears at larger $Ha$.
The contribution of the subgrid-scale dissipation and, thus,
the importance of the LES modeling, diminishes with the
strength of the magnetic field, but remains significant even at the strongest
field at which a continuous turbulent state is maintained.
 
\Thanks{ TB, DK and OZ acknowledge financial
support from the Deutsche Forschungsgemeinschaft (Emmy--Noether
grant Bo 1668/2-3 and Gerhard-Mercator visiting professorship
program). 
Computer resources were provided by the computing
centers of TU Ilmenau and the Forschungszentrum J\"ulich (NIC).
}

\newcommand{\noopsort}[1]{}


\lastpageno 


\end{document}